\documentclass[%
 reprint,
superscriptaddress,
 amsmath,amssymb,
 prc,
]{revtex4-2}

\usepackage{graphicx}
\usepackage{dcolumn}
\usepackage{bm}
\usepackage{CJKutf8}
\usepackage{color, lipsum}

\def\apjl{Astro.~Phys.~J.~Lett.}   
\def\nphysa{Nucl.~Phys.~A}   

\newcommand{\kanji}[1]{\begin{CJK}{UTF8}{ipxm}(#1)\end{CJK}}
\newcommand{\iso}[2]{{}^{#2}{\rm #1}}

\begin{document}

\preprint{}

\title{Postfission properties of uranium isotopes: A hybrid method with Langevin dynamics and the Hauser-Feshbach statistical model}

\author{S.~Tanaka \kanji{田中翔也}}
\email[]{shoya.tanaka@riken.jp}
\affiliation{RIKEN Nishina Center for Accelerator-Based Science, Wako, Saitama 351-0198, Japan}

\author{N.~Nishimura  \kanji{西村信哉}}
\email[]{nobuya.nishimura@riken.jp}
\affiliation{Astrophysical Big Bang Laboratory, RIKEN, Wako, Saitama 351-0198, Japan}
\affiliation{RIKEN Nishina Center for Accelerator-Based Science, Wako, Saitama 351-0198, Japan}

\author{F.~Minato \kanji{湊太志}}
\email[]{minato.futoshi@phys.kyushu-u.ac.jp}
\affiliation{Department of Physics, Kyushu University, Fukuoka 819-0395, Japan}
\affiliation{RIKEN Nishina Center for Accelerator-Based Science, Wako, Saitama 351-0198, Japan}

\author{Y.~Aritomo \kanji{有友嘉浩}}
\email[]{aritomo@ele.kindai.ac.jp}
\affiliation{Faculty of Science and Enginnering, Kindai University, Higashi-Osaka, Osaka 577-8502, Japan}

\date{\today}

\begin{abstract}
\begin{description}
\item[Background] Precise understanding of nuclear fission is crucial for experimental and theoretical nuclear physics, astrophysics, and industrial applications; however, the complete physical mechanism is unresolved due to the complexities.

\item[Purpose] In this study, we present a new method to describe the dynamical-fission process and following prompt-neutron emission, where we combine the dynamical fission calculation based on the Langevin method and the Hauser-Feshbach statistical model.

\item[Methods] Two methods are connected smoothly within the universal charge distribution and the energy conservation, allowing us to calculate a sequence of fission dynamics and postfission phase, including prompt neutron emission.

\item[Results] Using a certain set of model parameters, we successfully reproduce the experimental primary-fission yields, total kinetic energy, independent-fission yields, and prompt neutron emissions for the neutron-induced fission of ${}^{236}$U, a compound nucleus of ${\rm n} + {}^{235}{\rm U}$. We elucidate the physical mechanism of the characteristic features observed in previous experiments, such as shell properties. Additionally, we apply our calculation to two very neutron-rich uranium isotopes, i.e., ${}^{250}$U and ${}^{255}$U, which are not experimentally confirmed but are important for r-process nucleosynthesis. Theoretical results indicate that ${}^{250}$U exhibits an asymmetric multiple-peak fission yield distribution, while the neutron-rich ${}^{255}$U has a single peak due to symmetric fission. Our method predicts post-neutron emission fragments, where ${}^{250}$U shows a stronger neutron emissivity than ${}^{255}$U.

\item[Conclusions] Our framework is highly reproducible in the experiments and shows that the number of emitted neutrons after fission differs significantly in neutron-rich uranium fission depending on distributions of fission variables.

\end{description}
\end{abstract}


\maketitle

\section{Introduction}
\label{sec:Introduction}

Nuclear fission is a decay process in which a heavier nucleus is split into two or lighter nuclei, usually occurring in actinide elements (e.g., uranium) and further heavy nuclei~\cite{1939NW.....27...11H, 1939Natur.143..239M}. Fission can also be interpreted as the production process of unstable nuclei and nuclear-excited states, which may be accompanied by additional radioactive decay such as $\gamma$-ray decay and particle emission (e.g., neutrons and $\alpha$ particles). It is thus essential to understand the entire process, including the nuclear states of fission fragments involved in fission and subsequent particle emissions. A quantitative understanding of nuclear fission is crucial for nuclear engineering, but it is also an interesting subject in fundamental physics, such as nuclear physics and astrophysics. The fission of neutron-rich nuclei is of particular importance in understanding the physics of unstable nuclei and its application to r-process nucleosynthesis, which occurs in neutron-rich astrophysical environments, e.g., in compact-object-binary mergers \cite{2015ApJ...808...30E, 2018ApJ...863L..23Z, 2021PhRvC.103b5806L, 2022arXiv221204507W} (and see recent reviews, e.g, \cite{2020PrPNP.11203766A, 2021RvMP...93a5002C, 2022JPhG...49k0502S}).

The fission process begins with the dynamical behavior of unstable or excited nuclei, which is more challenging to describe within a theoretical model than a single nucleus in a stable state. Its study requires us to track the entire time evolution of a nucleus until it splits into multiple nuclei. Therefore, a full understanding of fission dynamics within ab initio approaches is still quite a difficult task. Many microscopic approaches based on energy density functionals have been examined \cite{Bender_2020, Schunck_2023, Zhao_2022, Zhao_2022a, Zhao_2022b, Guzman_2020, Taninah_2020, Bulgac_2022, Sadhukhan_2022, QiangYu_2021}, providing us plenty of knowledge to improve descriptions of fission dynamics. However, they are still not capable of reproducing experimental data of total kinetic energies (TKEs) and fragment yields simultaneously. Another approach to describe the fission process is the phenomenological method based on a fluctuation-dissipation method by Langevin equations \cite{2014PhRvC..90e4609A, PhysRevC.100.064605, PhysRevC.96.064616, ESLAMIZADEH2018163, PhysRevC.107.054616, PhysRevC.103.044601}, in which the dynamics are described by the motion of {\it classical} droplets, including quantum effects (such as nuclear-shell structure) as the nuclear potential. The method has successfully reproduced experimental values, including fission fragments and TKEs with appropriate physical parameters, and has been used not only in the field of applied nuclear physics but also in the research of the fundamental physics of unstable nuclei as a robust way to describe fission properties.

Fission is, however, a complex process where various phases are involved even after separations into multiple fragments. The primary-fission-product nuclei, usually in excited states, immediately undertake several decay processes, e.g., neutron emissions and radiative decays. The decay properties of actinides were measured precisely because of their critical importance in applications such as nuclear reactors. However, such a postfission process has not been considered in previous dynamical calculations and has yet to be compared with experimental data. Recently, the accuracy of TKEs and fission fragment yields calculated within the Langevin equation significantly improved, opening a new possibility to study radiative decays and fission dynamics comprehensively.

It is necessary to follow the postfission process precisely to explain experimental data on various fission. The postfission stage involves different physics from dynamical fission, so we must adopt other physical descriptions and models in addition to the Langevin approaches. To deal with particle evaporation from the postfission stage, the Hauser-Feshbach statistical model (HFSM) is one of the appropriate methods in which fission fragments are assumed to reach a thermal equilibrium immediately. The HFSM has already been applied to prompt neutron emissions for neutron- and photo-induced fission of actinide nuclei \cite{Okumura2018, Lovell2021, Okumura2021, FIFRELIN, Kawano_2023} as well as $\beta$-delayed neutron emissions \cite{Minato_2021, Mumpower_2022}. Comparison of theoretical calculations to various experiments enables the constrain and refinement of the adopted parameters in the Langevin model and HFSM.

In this study, we quantitatively investigate nuclear decay processes that occur subsequently to dynamical fission. The primary objective is establishing a calculation method that exhibits higher experimental reproducibility, particularly for significant physical phenomena such as neutron emissions. To this end, we incorporate the Langevin calculation [the Kindai University Langevin model (KiLM)] for fission dynamics and the HFSM for post-neutron emission (implemented in CCONE~\cite{2016NDS...131..259I}). We successfully reproduce experimental fission data by utilizing the hybrid method that connects the dynamical Langevin model and HFSM calculations. To study the performance of our framework, a neutron-induced reaction on $\iso{U}{236}$ is chosen. We also extend the framework to neutron-rich uranium isotopes of $\iso{U}{250}$ and $\iso{U}{255}$, which have yet to be experimentally evaluated.

The paper is structured as follows. Section~\ref{sec:methods} describes nuclear models and numerical methods for fission calculations and the postfission process. The results of $\iso{U}{236}$ and neutron-rich isotopes are shown in Section~\ref{sec:results}. Section~\ref{sec:summary} is devoted to summary and conclusions.

\section{Methods}
\label{sec:methods}

\begin{figure*}[htbp]
\includegraphics[width=0.9\hsize]{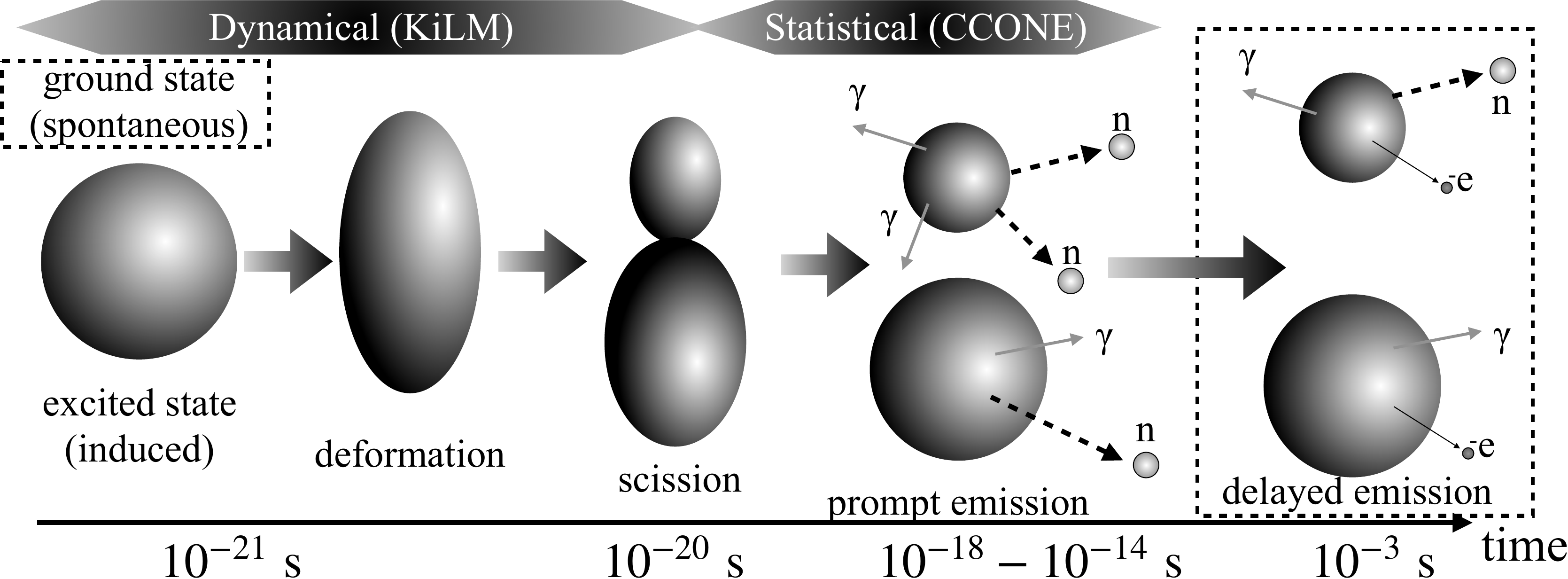}
\caption{\label{fig:schemetic_pic} Schematic representation of fission dynamics and the following prompt-neutron emission process (followed by delayed emission and decay). The change of nuclear shape splitting into two nuclei are described with typical time scales. See texts for details of the dynamical model by KiLM (Section~\ref{sec:dynamical_fiss}) and the statistical model by CCONE (Section~\ref{sec:statistical}).}
\end{figure*}

\begin{figure}[t]
\includegraphics[width=\hsize]{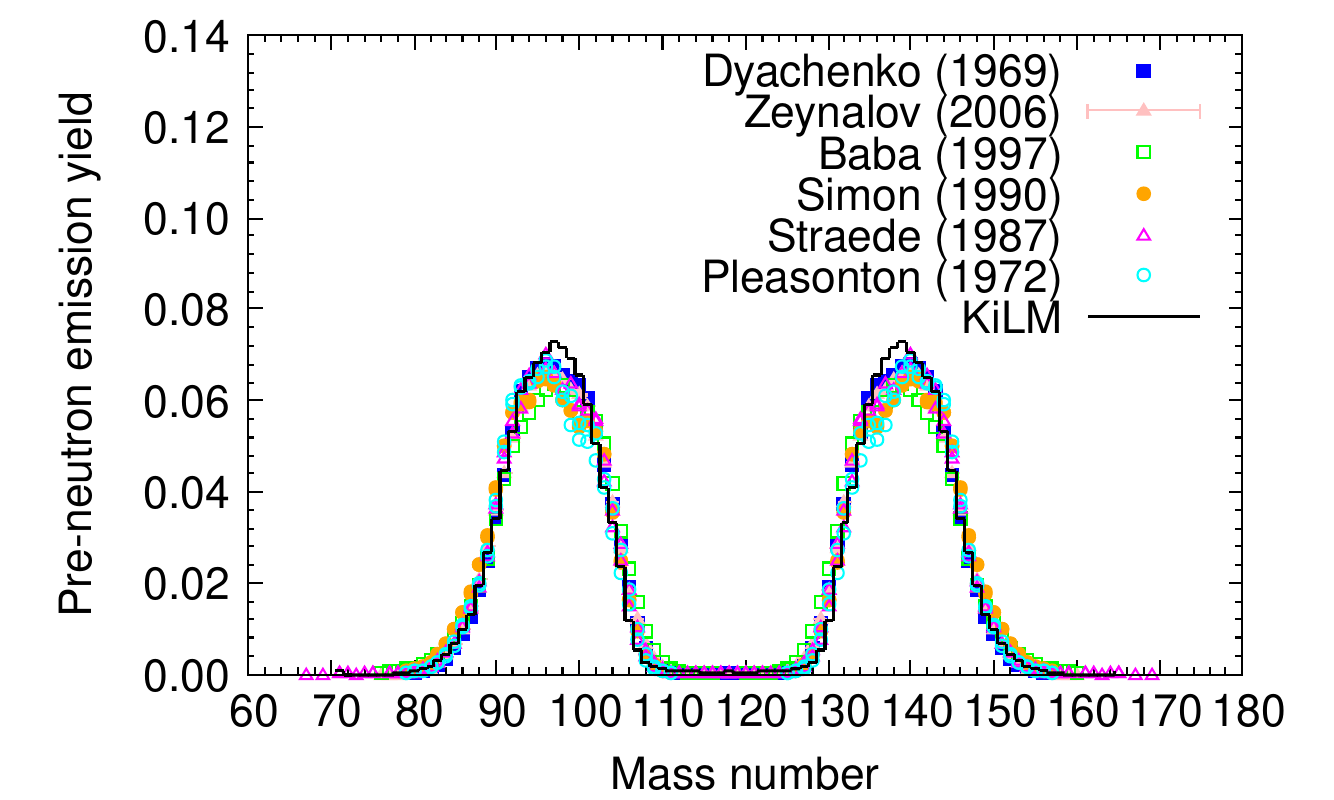}
\caption{\label{fig:u236_ffmd} The normalized fission yields of $\iso{U}{236}$, which is the compound nucleus of $\iso{U}{235} + {\rm n}$. The numerical results by KiLM are compared with several experimental data~\cite{dyachenko1969energy, Zeynalov2006, Baba_1997, SIMON1990220, STRAEDE198785, Pleasonton1972_PhysRevC.6.1023}.}
\end{figure}

To calculate the fission dynamics and the subsequent decay process of daughter nuclei, we employ a hybrid approach that combines the dynamical Langevin scheme and the HFSM. Our calculation scheme involves modeling a sequence of dynamical fission phases followed by postfission neutron emissions. Fig.~\ref{fig:schemetic_pic} shows a schematic picture of the fission process and prompt neutron emission that we investigate in this study. The following sections describe the calculation methods used in individual phases, including the physical parameter setups.

\subsection{The dynamical fission calculation}
\label{sec:dynamical_fiss}

To calculate nuclear-shape time evolution, we use the fluctuation-dissipation model with the Langevin equation, the KiLM \cite{2004NuPhA.744....3A, 2013PhRvC..88d4614A, 2022PhRvC.105c4604A}. The nuclear shape is defined by the two-center parametrization \cite{1972ZPhy..251..431M, 1978ZPhyA.288..383S}, which has three deformation parameters, $z$, $\delta$ and $\alpha$ to serve as collective coordinates, abbreviated as $q = \{ z, \delta, \alpha \}$. The symbol $z$ is the distance between two potential centers, the symbol $\delta$ denotes the deformation of the fragments, and ${\alpha = (A_{1} - A_{2})/(A_{1} + A_{2})}$ is the mass asymmetry of the two fragments \cite{2004NuPhA.744....3A}, where $A_{1}$ and $A_{2}$ denote the mass numbers of heavy and light fragments. Each fragment has the same degree of deformation ($\delta_1=\delta_2=\delta$).

For a given value of the temperature of a system $T$, which is related to the excitation energy of the composite system as $E^{*}=aT^{2}$ ($a$ is the level density parameter \cite{2014PhRvC..90e4609A, 1981NuPhA.372..141T}), the potential energy is defined as a sum of the liquid-drop (LD) part and a microscopic (SH) part:
\begin{equation}
\label{eq1}
\begin{split}
V (q,T)         &= V_{\rm LD}(q) + V_{\rm SH}(q,T),\\
V_{\rm LD}(q)   &= E_{\rm S}(q) + E_{\rm C}(q),\\
V_{\rm SH}(q,T) &= [\Delta E_{\rm shell}(q) + \Delta E_{\rm pair}(q) ] \Phi(T),\\
\Phi(T)         &= {\rm exp} \left( - \frac{aT^{2}}{E_{d}} \right ).
\end{split}
\end{equation}
Here, the potential energy $V_{\rm LD}$ is calculated with the finite-range liquid drop model \cite{1979PhRvC..20..992K}, given as a sum of the surface energy $E_{\rm S}$ and the Coulomb energy $E_{\rm C}$. The microscopic energy $V_{\rm SH}$ at $T\,=\,0$ is calculated as the sum of the shell correction energy $\Delta E_{\rm shell}$, evaluated by the Strutinsky method \cite{1967NuPhA..95..420S, 1968NuPhA.122....1S}, and the pairing correlation correction energy $\Delta E_{\rm pair}$ \cite{1967NuPhA..95..420S, 1969NuPhA.131....1N}. The shell correction energy has a temperature dependence expressed by a factor $\Phi$($T$) in which the shell damping energy $E_{d}$ is chosen as $20~{\rm MeV}$ \cite{ignatyuk1975fission}. We assume that the angular momentum of the fissioning nucleus is not large at the low excitation energy, so the orientation degree of freedom ($K$ coordinate) \cite{PhysRevC.85.064619}, and the rotational energy is not included in Eqs.~(\ref{eq1}).

To define the potential of the two-center shell model \cite{1972ZPhy..251..431M, 1978ZPhyA.288..383S}, a neck parameter of $\varepsilon = 0.35$ $(0 \le \varepsilon \le 1)$ \cite{1987NuPhA.475..487Y} has been routinely used \cite{2013PhRvC..88d4614A, 2014PhRvC..90e4609A, 2016PhLB..761..125L, 2016PhRvC..94d4602U, 2017PhRvL.119v2501H, 2017PhRvC..96f4617U}. However, this value is inappropriate for heavier actinide nuclides, as pointed out in \cite{2017PhRvL.119v2501H, 2019PhRvC..99e1601M}. We adopt the $\varepsilon$ values following an empirical formula \cite{2019PhRvC..99e1601M},
\begin{equation}
\varepsilon ( A_{c} ) = 0.01007 A_{c} - 1.94,
\label{eq2}
\end{equation}
where $A_{c}$ is the mass of the fissioning nucleus. This relation is adjusted to the overall trend of available experimental data of fission distributions for $Z \approx 92$--$98$ isotopes including $A>250$ (see the supplemental material of \cite{2019PhRvC..99e1601M}); we adopt $\varepsilon = 0.35, 0.58$, and $0.63$ for $\iso{U}{236}$, $\iso{U}{250}$, and $\iso{U}{255}$, respectively.

The multi-dimensional Langevin equations (see, e.g., \cite{2004NuPhA.744....3A}) are given as
\begin{equation}
\label{eq3}
\begin{split}
\frac{dq_{i}}{dt} =& \ (m^{-1})_{ij} p_{j},\\
\frac{dp_{i}}{dt} =& -\frac{\partial V}{\partial q_{i}} - \frac{1}{2} \frac{\partial}{\partial q_{i}} (m^{-1})_{jk} p_{j} p_{k}\\
& \qquad - \gamma_{ij} (m^{-1})_{jk} p_{k} + g_{ij} R_{j} (t),
\end{split}
\end{equation}
where $q_{i}$\,=\,$\{ z, \delta, \alpha \}$ and $p_{i}$\,=\,$m_{ij} dq_{i}/dt$ is a momentum conjugate to coordinate $q_{i}$. In the Langevin equation, $m_{ij}$ and $\gamma_{ij}$ are the shape-dependent collective inertia and the friction tensors, respectively. The wall-and-window one-body dissipation \cite{1978AnPhy.113..330B, 1984NuPhA.428..161R, 1987RPPh...50..915F} is adopted for the friction tensor.
A hydrodynamical inertia tensor is calculated with the Werner-Wheeler approximation for the velocity field \cite{PhysRevC.13.2385}.

The normalized random force $R_{i} (t)$ is assumed to be that of white noise, i.e.,
\begin{align}
\label{eq4}
\langle R_{i} (t) \rangle &= 0,\notag\\
\langle R_{i} (t_{i}) R_{j} (t_{2}) \rangle &= 2 \delta _{ij} \delta (t_{1} - t_{2}).
\end{align}
The strength of the random force $g_{ij}$ is related to the friction tensor $\gamma_{ij}$ by the classical Einstein relation \cite{Hoffman1998},
\begin{align}
\label{eq5}
\sum_{k} g_{ik} g_{jk} = \gamma_{ij} T^{*}, \notag\\
T^{*} = \frac{\hbar\omega}{2} \coth \left(\frac{\hbar\omega}{2T} \right),
\end{align}
where $T^*$ is the effective temperature \cite{PhysRevC.96.064616}. The parameter $\omega$ is the local frequency of collective motion~\cite{Hoffman1998}. The minimum of $T^*$ is given by $\hbar\omega\slash2$, which corresponds to the zero-point energy of oscillators forming the heat bath. We estimated the zero-point energy as $3.05$~MeV to be consistent with experimental data, which is higher than the previous study.

The random properties introduced in Eqs.~(\ref{eq3})--(\ref{eq5}) give different trajectories on the potential energy space event-by-event, creating the fission fragment mass distribution by accumulating enough number of different trajectories, which can be directly compared to the experimental data. In each trajectory, fission is defined as the case that nucleus reaches the scission point on the potential energy surface. The calculation result for fission fragment mass distribution of $\iso{U}{236}$ ($E^{*}=9$~MeV) is shown in Fig.~\ref{fig:u236_ffmd} with several experimental data, where the mass distribution is reasonably reproduced. This excitation energy is adopted for numerical efficiency, which may be relatively higher than experiments. However, it is a lower value than ones adopted in previous Langevin calculations \cite{2013PhRvC..88d4614A, 2014PhRvC..90e4609A} and the threshold of second-chance fission \cite{PhysRevC.100.064605,2017PhRvL.119v2501H}. 

\subsection{The unchanged-charge distribution and total kinetic energy (TKE)}
\label{sec:ucd}

Since the dynamical fission calculations with the KiLM consider the mass of each nucleus but do not the atomic number ($Z$), we assume the {\it unchanged-charge distribution} (UCD) hypothesis to estimate fission fragment charge distributions. The UCD assumption is a simple treatment; the most probable charge of a fission fragment $Z_{\rm{f}}$ is determined from the ratio of an atomic number to mass of parent nuclei, i.e., $Z_{\rm f}=A_{\rm f}Z_{\rm CN}/A_{\rm CN}$, where $A_{\rm CN}$ and $Z_{\rm CN}$ are mass and atomic number of compound nucleus, and $A_{\rm f}$ is the fragment mass, respectively.

\begin{figure}[t]
\includegraphics[width=\hsize]{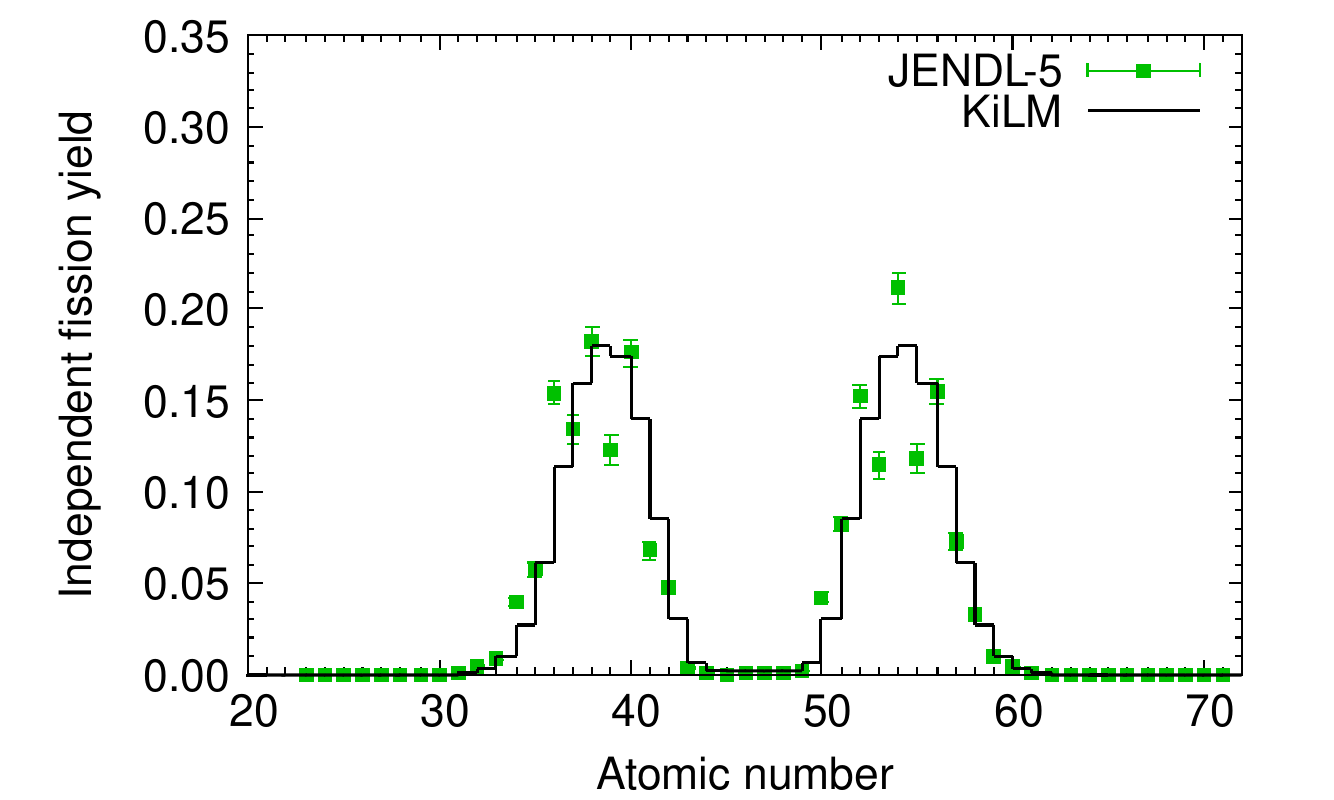}
\caption{\label{fig:u236_IAY} The normalized fission yield distribution of $\iso{U}{236}$ against the atomic number (the charge distribution), based on the UCD. The results of the KiLM are compared with the evaluated data of JENDL-5 \cite{JENDL-5}.}
\end{figure}

Experimental data show that the charge distribution of fission fragments slightly deviates from the UCD \cite{Wahl2002, ER1993}; however, it is not easy to estimate the deviations for neutron-rich nuclei where no experimental data are available. We, therefore, use the UCD assumption mentioned above to estimate charge distributions in our calculations. Fortunately, the deviations from UCD do not significantly affect prompt neutron yields. However, as pointed out in Refs.~\cite{Minato_2018, Okumura2021}, we need consider to the charge distribution more seriously if delayed neutron yields are discussed. We want to mention one theoretical approach that has calculated charge fragmentation of $\iso{Pu}{240}$ within a Hartree-Fock-Bogoliubov + particle number projection \cite{PhysRevC.100.024612}, which has the potential to predict charge distributions of unmeasured fissioning nuclei.

To consider the charge polarization, the fragment charge number $Z_1$ is randomly determined by following a Gaussian distribution based on the UCD assumption in KiLM, where the mean value $\mu_{Z_{1}}$ is set to
\begin{equation}
    \mu_{Z_{1}} = \frac{Z_{\rm CN}}{A_{\rm CN}}A_{\rm 1}-0.5
\end{equation}
with the standard deviation of $\sigma = 0.493$, a typical value of the charge distribution~\cite{ER1993,WAHL19881}. The $Z_2$ is automatically determined by $Z_{2}=Z_{\rm CN}-Z_{1}$. The calculated charge distribution of $\iso{U}{236}$ ($E^{*}=9$~MeV) is shown in Fig.~\ref{fig:u236_IAY}. Although an odd-even staggering is observed in the evaluated data of charge distribution of ${\rm n_{th}}+\iso{U}{235}$~\cite{JENDL-5}, our calculation reasonably reproduces the distribution.

\begin{figure}[t]
\includegraphics[width=\hsize]{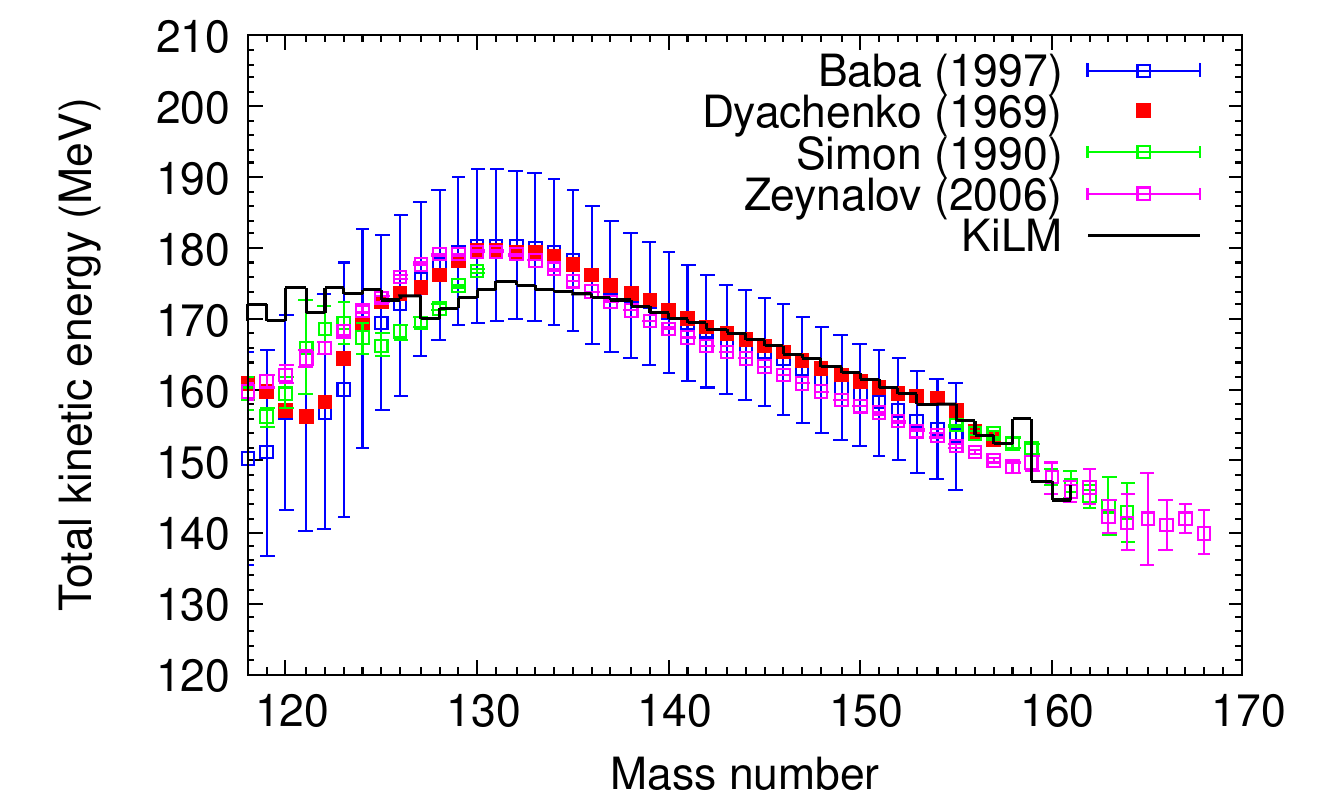}
\caption{\label{fig:u236_TKE} The mean value of the TKE for $\iso{U}{236}$. The results by KiLM are compared with experimental data \cite{Baba_1997, dyachenko1969energy, SIMON1990220, Zeynalov2006}.}
\end{figure}

The TKE of fission fragments is assumed to be given by
\begin{equation}
    {\rm TKE} = V_{\rm Coul}+E_{\rm pre},
\end{equation}
where $V_{\rm Coul}=Z_1Z_2e^2/D$ and $E_{\rm pre}$ are the Coulomb repulsion energy of point changes of fragments and the pre-scission kinetic energy. The distance between the centers of mass of the left and right parts of the nucleus at the scission point is represented by $D$. The pre-scission kinetic energy $E_{\rm pre}$ is given by
\begin{equation}
    E_{\rm pre} = \frac{1}{2}(m^{-1})_{ij}p_ip_j,
\end{equation}
which is calculated at the scission point. The statistical average of $E_{\rm pre}$ of $\iso{U}{236}$ ($E^{*}=9$~MeV) in the KiLM is equal to $5.93$~MeV. Therefore, the main contribution to the TKE comes from the Coulomb repulsion of fission fragments. The mean value of the TKE ($\langle \mathrm{TKE} \rangle$) as a function of the fission fragment mass is shown in Fig.~\ref{fig:u236_TKE} for the fission of $\iso{U}{236}$ at $E^*=9$~MeV. Averaging over fission mass yields, we obtain $\langle \mathrm{TKE} \rangle = 170.16~{\rm MeV}$. The KiLM overestimates the $\langle \mathrm{TKE} \rangle$ around $A = 120$ and underestimates around $A = 130$. On the other hand, it reproduces well for $A \ge 136$. We expect the deviations of $\langle \mathrm{TKE} \rangle$ from $A = 120$ to $136$ does not affect too much prompt-neutron emissions because the fragment mass yields are much smaller than $A \ge 136$ as seen in Fig.~\ref{fig:u236_ffmd}.

\subsection{The Hauser-Feshbach statistical model}
\label{sec:statistical}

We use the HFSM with width-fluctuation correction implemented in CCONE code~\cite{2016NDS...131..259I} to estimate particle evaporation from excited fission fragments at the postfission evolution, including the prompt neutron emission. CCONE is one of the widely used codes that comprehensively covers various nuclear reactions. It has been applied to produce proton-, neutron-, deuteron-, and photo-nuclear reactions in JENDL-5 \cite{JENDL-5} and particle evaporation following $\beta$-decay \cite{Minato2021} and muon captures \cite{Minato2023}, giving a good agreement with experimental data. To carry out the HFSM calculation, excitation energy and spin-parity distributions of fragments defined as $\rho(J^{\pi}, E^{*})=\rho(J^{\pi})\rho(E^{*})$ are needed. The energy partition between light and heavy fragments is determined by an anisothermal parameter $R_{T}$ defined by~\cite{KAWANO201351}
\begin{equation}
R_{T}=\frac{T_{l}}{T_{h}}=\sqrt{\frac{a_{h}(U_{h})U_{l}}{a_{l}(U_{l})U_{h}}},
\label{eq:anisothermal}
\end{equation}
where $a_{l,h}$ is the energy-dependent level density parameter and $U_{l,h}$ is the energy corrected by the pairing energy $\Delta$, $U_{l,h}=E_{l,h}-\Delta_{l,h}$. The excitation energy $E_{l,h}$ are determined in an iterative way from Eq.~\eqref{eq:anisothermal}. Here, we define the total excitation energy (TXE) as
\begin{equation}
\mathrm{TXE}=E_{l}+E_{h}=M_{C}-(M_{l}+M_{h})-\mathrm{TKE}+E^{*},
\end{equation}
where $M_{C}$, $M_{l},$ and $M_{l}$ are the mass of the compound system, light fragment, and heavy fragment, respectively. We set in this work $R_{T}=1.2$ that is determined from the n+$^{235}\mathrm{U}$ reaction~\cite{Lovell2021}. The excitation energy distribution is then calculated by~\cite{Okumura2018,Okumura2021,Lovell2021}
\begin{equation}
\rho(E^{*}_{l,h})=\frac{1}{\sqrt{2\pi\delta_{l,h}^2}}
\exp\left(-\frac{(E^{*}_{l,h}-E_{l,h})^{2}}{2\delta_{l,h}^{2}}\right),
\end{equation}
where the width parameter $\delta_{l,h}$ is estimated by
\begin{equation}
    \delta_{l,h}=\frac{\delta_{\mathrm{TXE}}}{\sqrt{E_{l}^{2}+E_{h}^{2}}}E_{l,h}.
\end{equation}
The width parameter $\delta_{\mathrm{TXE}}=\delta_{\mathrm{TKE}}$ is calculated from KiLM and UCD described in Sect.~\ref{sec:dynamical_fiss} and \ref{sec:ucd}.

For the spin-parity distribution of fragments, a widely-applied expression:
\begin{equation}
\rho(J^{\pi})=\frac{1}{2}\frac{J+1/2}{2(f\sigma(U))^{2}}\exp\left(-\frac{(J+1/2)^2}{2(f\sigma(U))^{2}}\right),
\end{equation}
is used where $\sigma(U)$ is the spin cutoff parameter, and we use the same value of $f=2.756$ as determined from the n+$^{235}$U reaction~\cite{Lovell2021}.

Transmission coefficients of nucleons are calculated by the optical potentials of Koning-Delaroche~\cite{KandD2003}. For nuclear level densities, the Gilbert-Cameron method~\cite{GC} with Mengoni-Nakajima parameter~\cite{MandN} is adopted. For $\gamma$ strength functions, the enhanced generalized Lorentzian function~\cite{KandU} is used. Mass data are taken from the AME2020~\cite{Huang2021}.


\begin{figure}[t]
\includegraphics[width=\hsize]{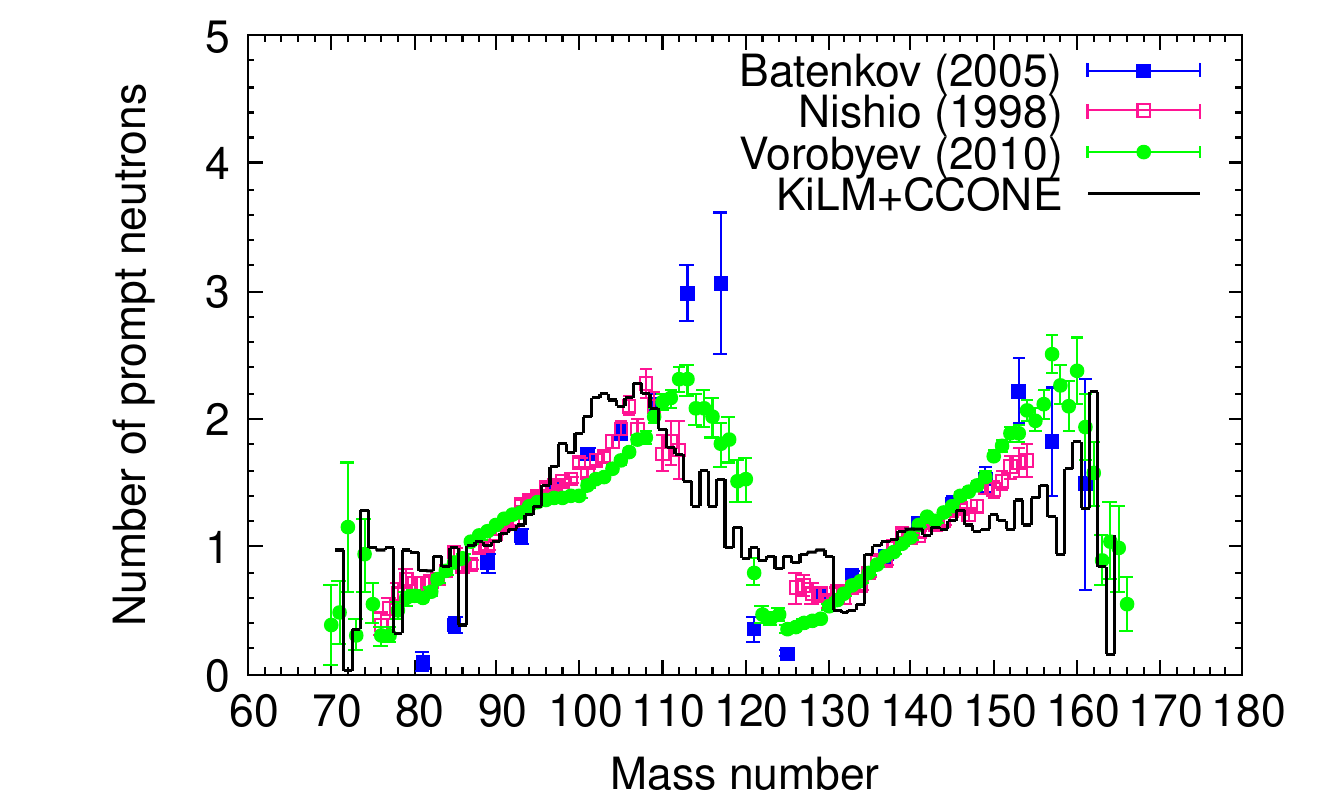}
\caption{\label{fig:prompt_n_u236} The number of prompt neutrons for $\iso{U}{236}$. The calculated values with a solid line (KiLM+CCONE) as a function of the mass number are compared with evaluated data \cite{Batenkov:2005uns, NISHIO1998540, Vorobyev2010}. The calculated average neutron emission number is $\langle \mathrm{n} \rangle = 2.413$.}
\end{figure}

\begin{figure}[t]
\includegraphics[width=\hsize]{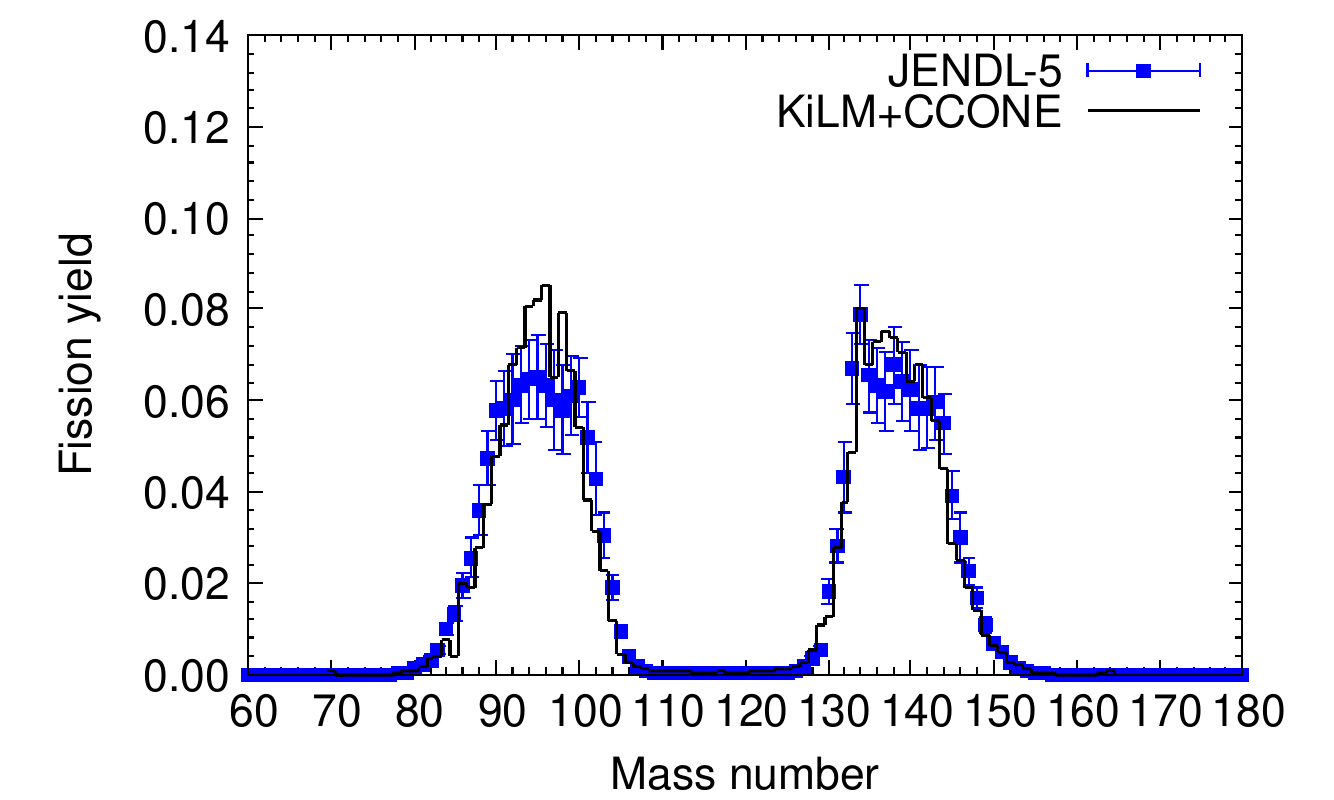}
\caption{\label{fig:u236_ffmd_post} The normalized fission yields of $\iso{U}{236}$ (the compound nucleus of $\iso{U}{235} + {\rm n}$) calculated by KiLM+CCONE after the prompt neutron emission is compared with evaluated data (JENDL-5 \cite{JENDL-5}).}
\end{figure}

\begin{figure*}[t]
\includegraphics[width=0.45\hsize]{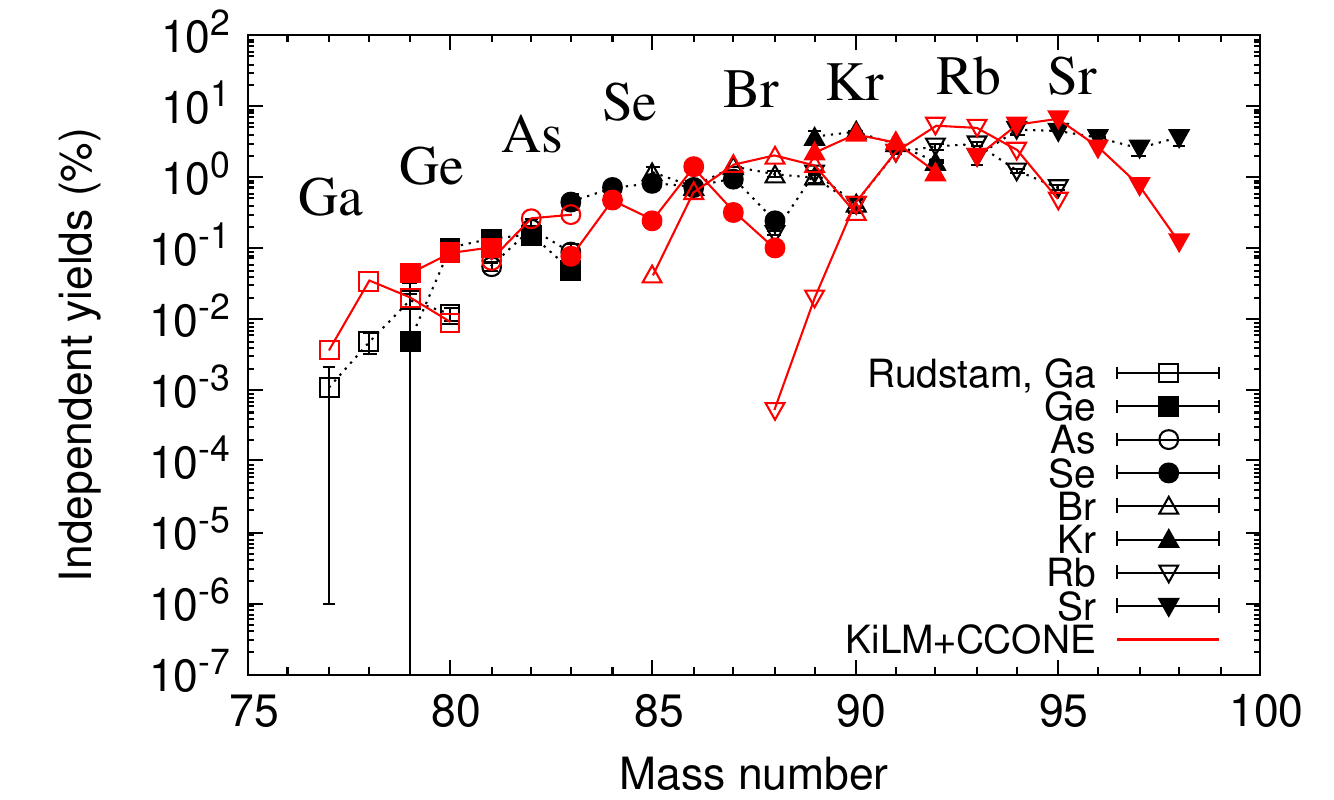}
\includegraphics[width=0.45\hsize]{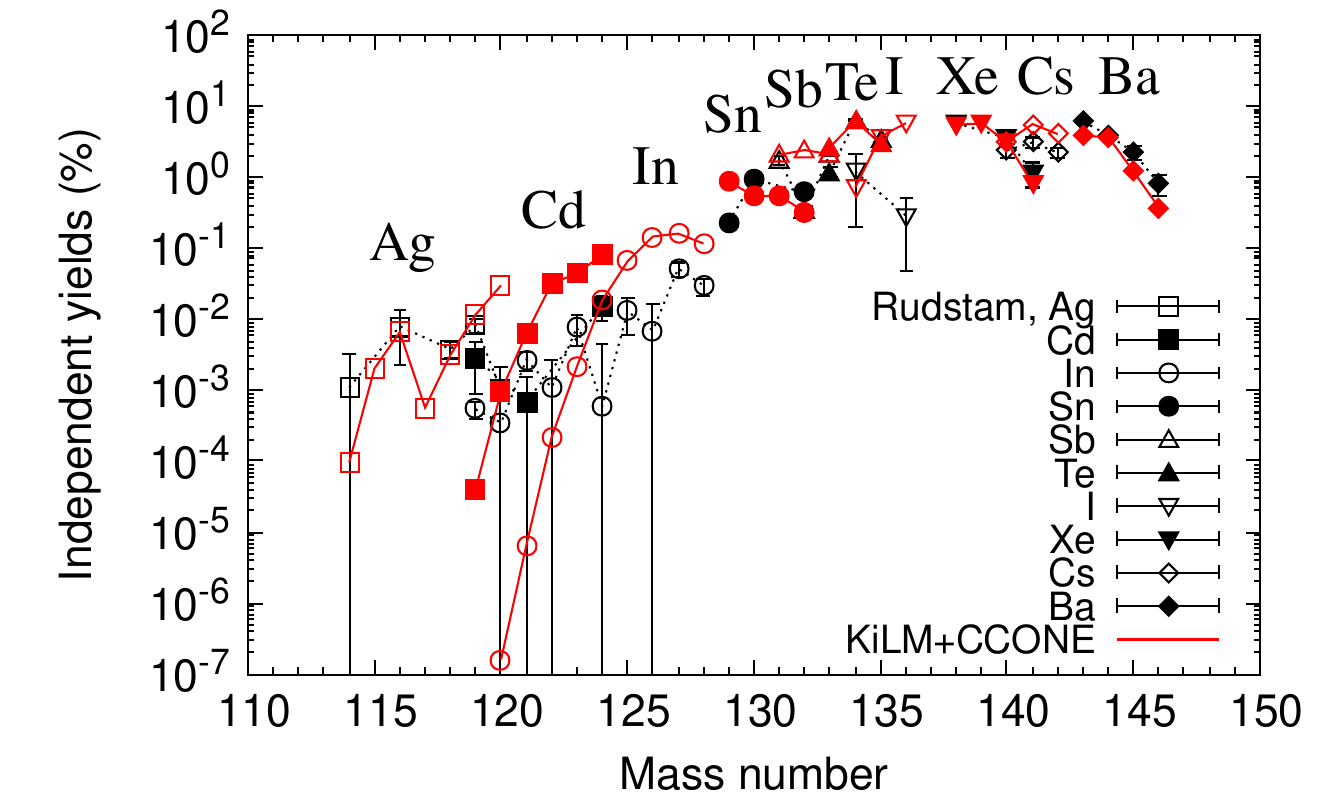}
\caption{\label{fig:u236_ffd_iso} The independent yields of individual isotopes with experimental data \cite{1990ADNDT..45..239R}. Experimentally identified isotopes are selected: $\iso{Ga}{77-80}$, $\iso{Ge}{79-83}$, $\iso{As}{81-83}$, $\iso{Se}{83-88}$, $\iso{Br}{85-90}$, $\iso{Kr}{89-92}$, $\iso{Rb}{88-95}$, and $\iso{Sr}{93-98}$ in the left panel and $\iso{Ag}{114,116,118-120}$, $\iso{Cd}{119,120,124}$, $\iso{In}{119-128}$, $\iso{Sn}{129,130,132}$, $\iso{Sb}{131-133}$, $\iso{Te}{133-135}$, $\iso{I}{134,136}$, $\iso{Xe}{138,140,141}$, $\iso{Cs}{140-142}$, and $\iso{Ba}{143-146}$ in the right panel.}
\end{figure*}

\begin{figure}[h]
\includegraphics[width=\hsize]{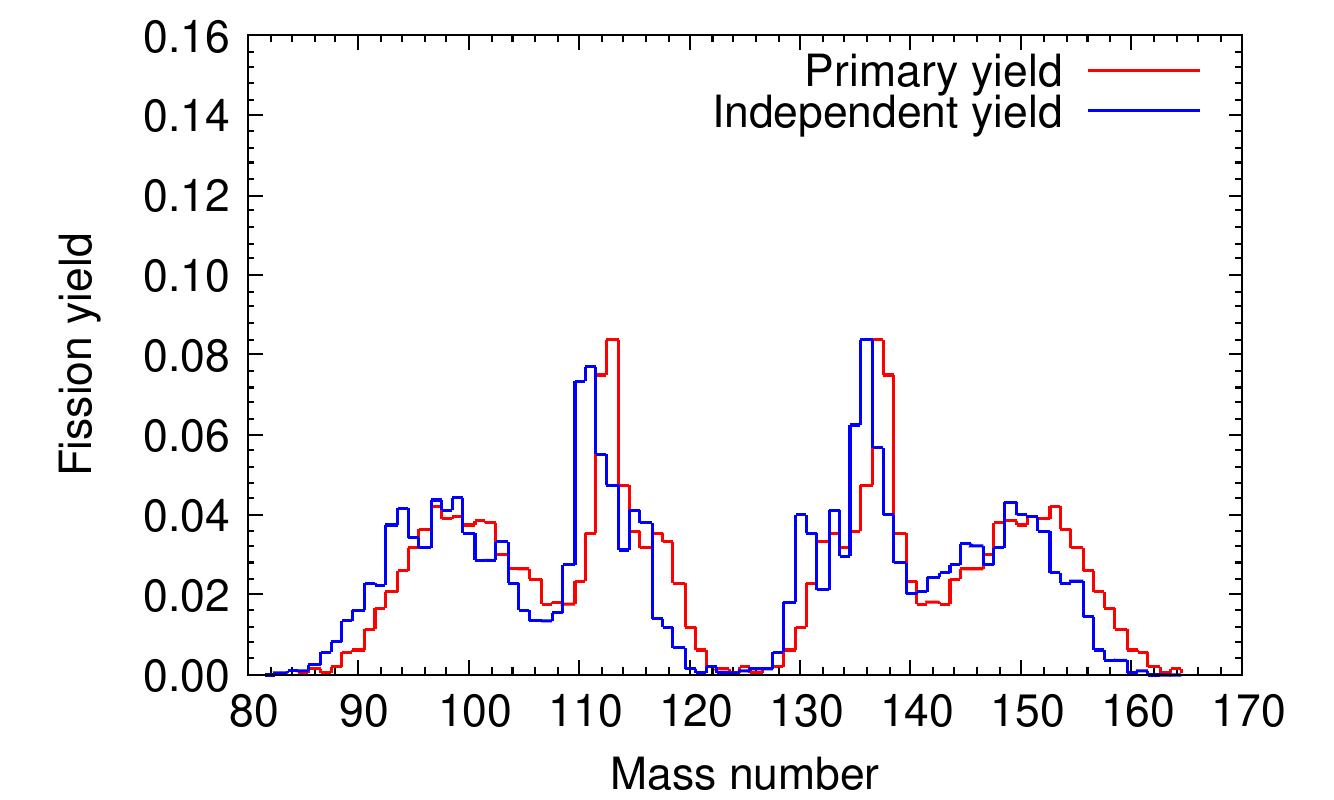}
\vspace{10pt}
\includegraphics[width=\hsize]{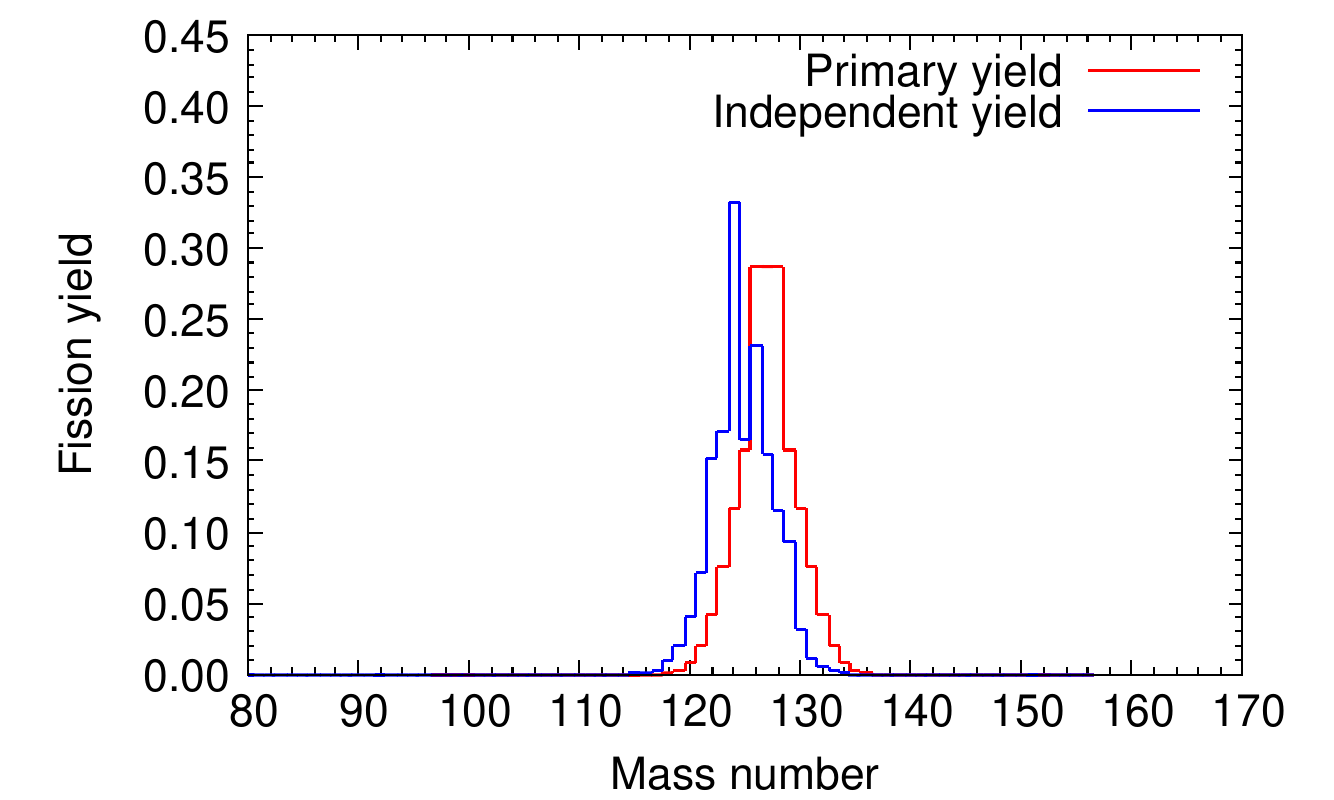}
\caption{\label{fig:ffmd_u_nrich} The normalized distribution of fission yields for $\iso{U}{250}$ (top) and $\iso{U}{255}$ (bottom). The primary yield distribution and the independent fission yields are compared.}
\end{figure}

\begin{figure}[h]
\includegraphics[width=\hsize]{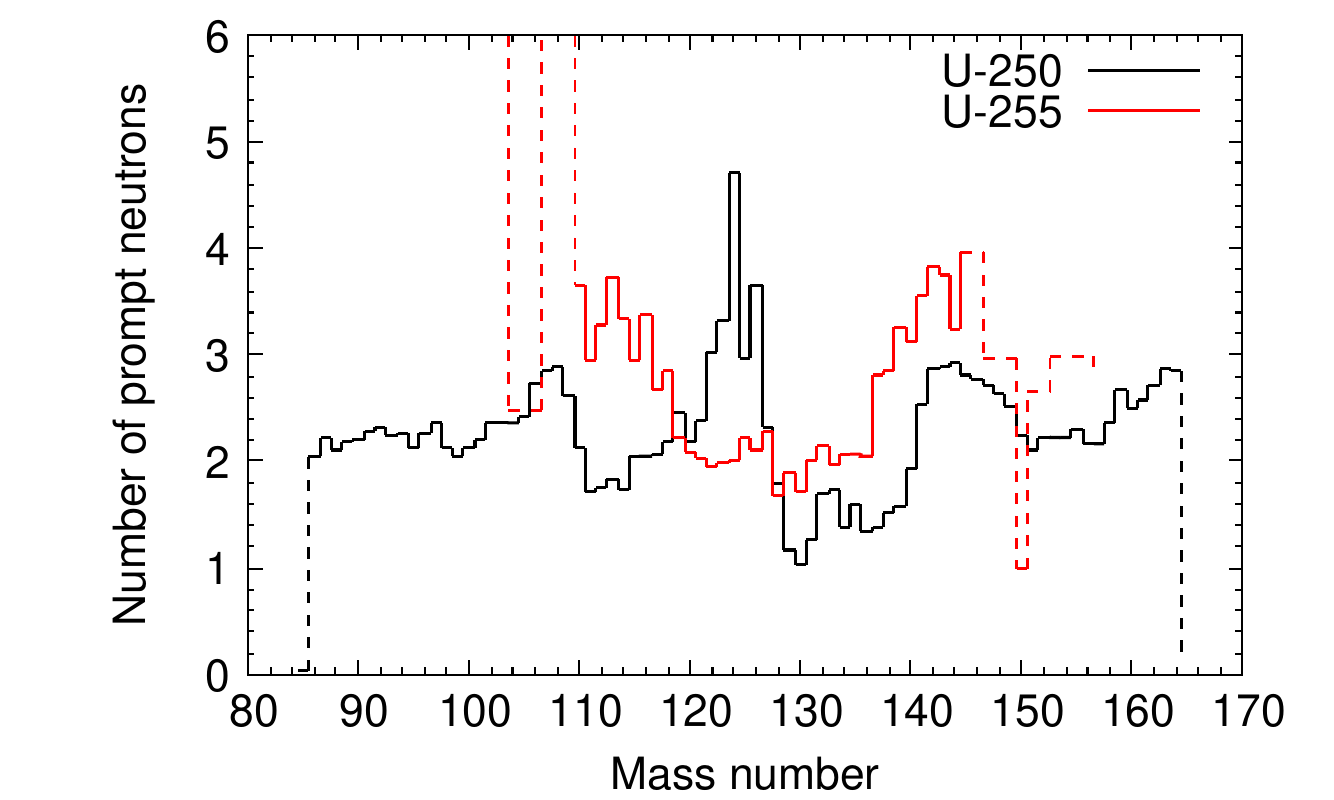}
\caption{\label{fig:prompt_n_u_nrich} The number of prompt neutrons as a function of fragment mass for $\iso{U}{250}$ (black line) and $\iso{U}{255}$ (red line). The solid line corresponds the statistically significant range, while the dotted indicate lower statistics influenced by one or few fission events. The average neutron emission numbers are $\langle \mathrm{n} \rangle = 4.185$ for $\iso{U}{250}$ and $\langle \mathrm{n} \rangle = 3.434$ for $\iso{U}{255}$.}
\end{figure}

\begin{figure}[h]
\includegraphics[width=\hsize]{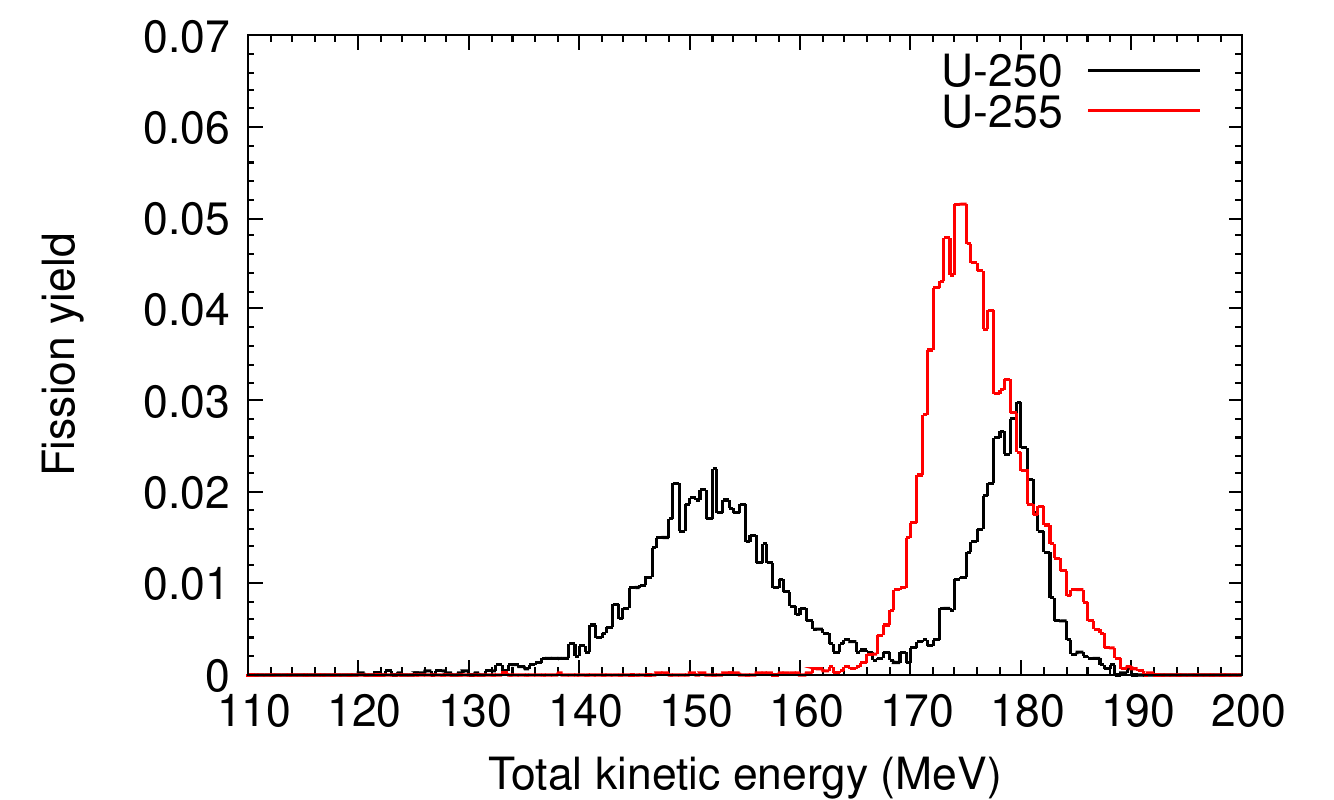}
\caption{\label{fig:tke_nrich} The fission yield distribution in the TKE for $\iso{U}{250}$ (black line) and $\iso{U}{255}$ (red line). The average TKE values are $\langle \mathrm{TKE} \rangle = 162.50~{\rm MeV}$  for $\iso{U}{250}$ and $\langle \mathrm{TKE} \rangle = 175.85~{\rm MeV}$ for $\iso{U}{255}$.}
\end{figure}

\begin{figure}[t]
\includegraphics[width=\hsize]{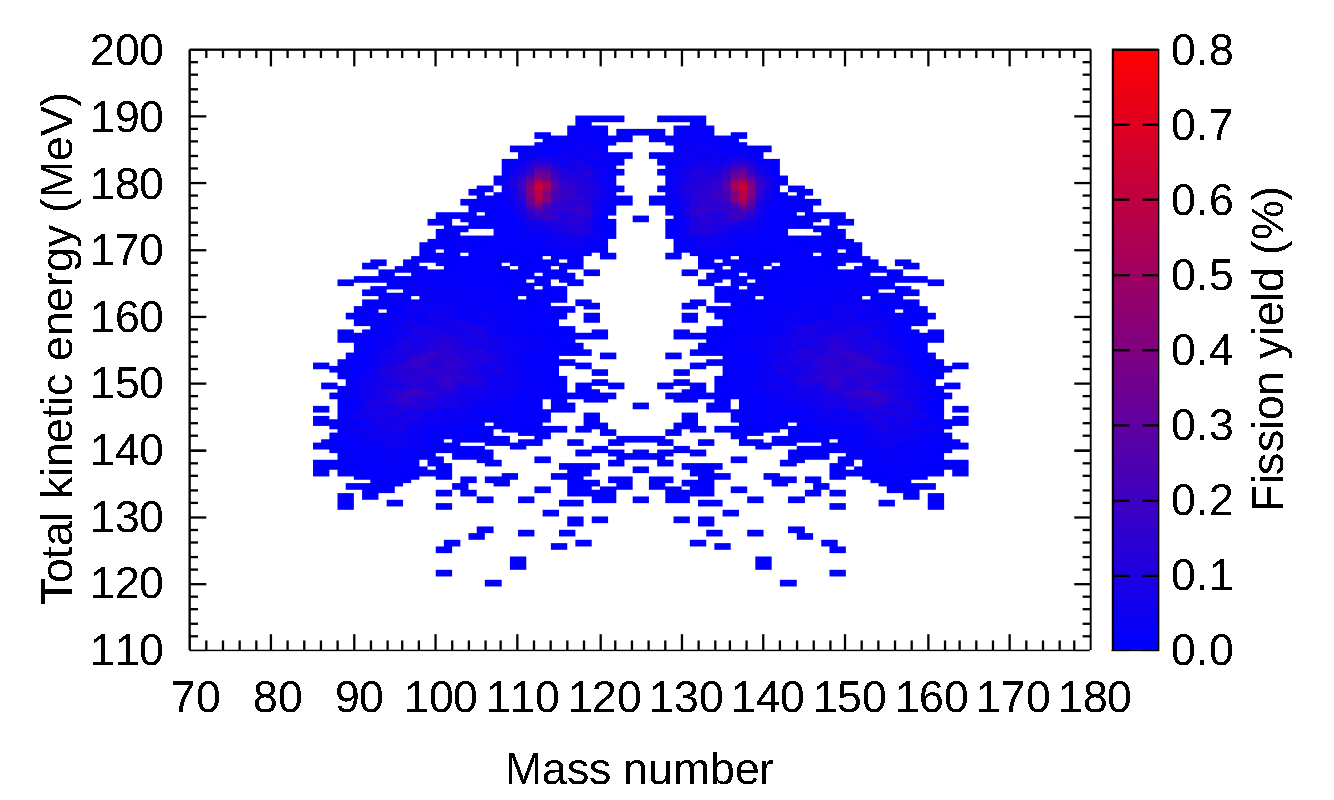}
\caption{\label{fig:tke_mass_u250} The fission yield distribution of $\iso{U}{250}$ on the TKE--mass number plane. The color scale indicates fission yields.}
\end{figure}

\section{Results}\label{sec:results}

\subsection{Fission properties of $\iso{U}{236}$}

As described in Section~\ref{sec:methods}, we calculate the prompt neutron emission process within the HFSM implemented in CCONE, following the dynamical calculation with the KiLM for the induced fission of $\iso{U}{236}$, i.e., a compound nucleus of ${\rm n}+\iso{U}{235}$ in the present study. The outputs calculated by KiLM for $\iso{U}{236}$, e.g., the mass distribution of fission yields (Fig.~\ref{fig:u236_ffmd}), the charge distribution with the UCD assumption (Fig.~\ref{fig:u236_IAY}), and the TKE distributions (Fig.~\ref{fig:u236_TKE}) are collected to carry out the calculation of CCONE. 

The numerical results of the number of prompt neutrons as a function of the mass number are shown in Fig.~\ref{fig:prompt_n_u236}. The global trend of experiments, the so-called saw-tooth structure that yields the increase from $A = 70~(130)$ to around $110~(160)$ and the decrease from $A=110~(160)$ to $120~(170)$, are reproduced feasibly although the position of the peaks deviates from the experimental data. Although the physical origin of the saw-tooth structure is not clarified well, we can qualitatively understand that suppression around $A=120$ is due to the shell effect of Sn isotopes with $Z = 50$, where its decay and particle emission are weakened. The broad framework is thus characterized by the strong influence of dynamics and shell potentials. Our calculated results are close to the experimental data by Nishio~(1998) \cite{NISHIO1998540} among available experimental data but have discrepancies with the other two experiments near the lower boundary ($A < 80$) and higher boundary ($A > 150$), as well as a steep drop in the range of $A \approx 110$ to $120$. Because experimental results of those regions still exhibit significant dispersion, further investigation in terms of both experiment and theoretical models must be necessary.

Fig.~\ref{fig:u236_ffmd_post} shows the independent fission yields after prompt-neutron emission. The evaluated data by JENDL-5 is compared with our numerical results by the KiLM+CCONE calculation. The distribution represents the independent fission yields after neutron emission from the primary fission shown in Fig.~\ref{fig:u236_ffmd}. The corresponding fission yield reduces by a few mass units due to few neutron emissions. We can see that heavier fission yields in the range of $A = 130$ to $150$, in particular, exhibit excellent agreement with evaluated data. This indicates the high reproducibility of our fission+neutron emission calculations. On the other hand, while the overall distribution of lighter peaks is well reproduced, there are some discrepancies in areas that exhibit complex structures. Based on the calculation, the average number of the prompt neutrons per one fission event is $\langle {\rm n} \rangle = 2.574$, which is in good agreement with the evaluated value of $\langle {\rm n} \rangle = 2.413$~\cite{JENDL-5}. Although there are some minor discrepancies in the fission yield distribution, the overall features are well reproduced.

Fig.~\ref{fig:u236_ffd_iso} presents the isotopic distributions of fission products, compared with the experimental data~\cite{1990ADNDT..45..239R}. As expected by the high reproducibility of fission yields, our results reproduce experiments well, except for nuclei with large experimental errors or limited statistics in fission yields. This highlights the robustness of our calculations across multiple data sets, including charge distributions. We should note, however, that our dynamics calculations still utilize a simplified charge distribution with the UCD assumption, which may cause some discrepancies. Further investigations with advanced treatment of the charge distribution~(e.g., \cite{PhysRevC.100.024612, 2023EPJWC.28404008E}) are still expected to make our framework more sophisticated.

\subsection{Neutron-rich U isotopes: $\iso{U}{250}$ and $\iso{U}{255}$}

Using the KiLM+CCONE method, we calculate nuclear fission and prompt-neutron emissions for two additional neutron-rich U isotopes, $\iso{U}{250}$ and $\iso{U}{255}$ with the same excitation energy as $\iso{U}{236}$  ($E^{*}=9$~MeV). The calculated primary fission yields before prompt neutron emission and independent fission yields after prompt neutron emission are shown in Fig.~\ref{fig:ffmd_u_nrich}. The yields of $\iso{U}{250}$ show four peaks (two strong and moderate double peaks) due to two different mass-asymmetric fission modes, while those of $\iso{U}{255}$ have a single peak by mass-symmetric fission. Our previous dynamical fission calculations also show the transition from asymmetric to symmetric fission in neutron-rich U isotopes, the preliminary results of which were reported in Refs.~\cite{2021arXiv210509272O, 2023EPJWC.27911021T}. A similar transition has also been experimentally suggested in Fm isotopes with mass numbers close to the present study ($\iso{Fm}{254,256,258}$) \cite{PhysRevC.16.1483, PhysRevC.5.1725, PhysRevC.40.770} and a wide range of other elements (see, e.g., \cite{2015NuPhA.944..204I} for a review).

Here, we focus on how the postfission behavior changes when the nucleus becomes neutron-rich, and fission yield distributions become symmetric from asymmetric. The fission theory of neutron-rich nuclei remains unresolved and must continue to be investigated theoretically and experimentally. Fig.~\ref{fig:ffmd_u_nrich} shows that the independent fission yields distribute in smaller mass regions than the primary fission due to the prompt neutron emissions. The characteristic point of the independent fission yields is a more detailed distribution than the primary fission yields, which originates from the shell structure of fission fragments. For example, staggering around $A\approx 96, 130$ become prominent for $\iso{U}{250}$ and the independent yields show two peaks at $A=124$ and $126$ for $\iso{U}{255}$ that is not present in the primary yields.

The number of emitted neutrons as a function of fragment mass for $\iso{U}{250}$ and $\iso{U}{255}$ is plotted in Fig.~\ref{fig:prompt_n_u_nrich}. The solid line represents the statistically significant range, while the dotted line may overestimate the average emitted neutrons with tiny fission products, which have negligible impact on the overall discussion. The results of those neutron-rich nuclei are different from $\iso{U}{236}$ in Fig.~\ref{fig:prompt_n_u236}. A significant difference is the disappearance of the saw-tooth structure observed for $\iso{U}{236}$. Instead, a peak is found in $A \approx 125$ for $\iso{U}{250}$, and a large number of neutrons is emitted from light fragments and $A \approx 144$ for $\iso{U}{255}$. Analyzing the calculated results, the peak for $\iso{U}{250}$ and many neutron emissions for $\iso{U}{255}$ resulted from the contributions from many fission fragments rather than those from one or a few specific nuclei. In other words, as discussed in the following paragraph, this is relevant to the excitation energy of fission fragments. Those outstanding peaks do not contribute significantly to the average number of prompt neutrons because the corresponding fission yields are relatively small.

Fig.~\ref{fig:tke_nrich} is the case of $\iso{U}{250}$, two distinct peaks are seen at $\mathrm{ TKE} \approx 150~{\rm MeV}$ and $\approx 180~{\rm MeV}$. In contrast, $\iso{U}{255}$ exhibits only a single peak around $\mathrm{TKE} =175~{\rm MeV}$. The single peak structure is a natural consequence because $\iso{U}{255}$ does mass-symmetric fission, while two peak structure found in $\iso{U}{250}$ originates from the fact that $\iso{U}{250}$ does asymmetric fission having four peak structure in the mass distributions as seen in Fig.~\ref{fig:ffmd_u_nrich}. Considering the energy conservation, fragment pairs with a small TKE have a large TXE, while those with a high TKE have a small TXE. Namely, most fission fragments with $A \approx 115~(135)$ for the induced-fission of $\iso{U}{250}$ have a rather larger TXE than those for $\iso{U}{255}$. To explain this, we also plot in Fig.~\ref{fig:tke_mass_u250} the fission yield distribution of $\iso{U}{250}$ on the TKE--mass number plane. We can identify that $\mathrm{TKE}=150~{\rm MeV}$ and $180~{\rm MeV}$ peaks of $\iso{U}{250}$ correspond to the peak-pairs of $A \approx 115$ (and 135) and $A \approx 95$ (and 150), respectively. Since the Coulomb energies at the scission point ($V_{\rm Coul}$), the major contributor to TKE, are different for fragment pairs, two peaks of TKE can be explained by four peak structures in the mass distributions. We can also see that the distribution of TKE around $A \approx 124$ relatively concentrates on smaller TKE around $130$--$140$~MeV, resulting in fission fragments with high excitation energies. As a consequence, we have a sharp peak around $A \approx 124$ for the number of prompt neutrons of $\iso{U}{250}$ in Fig.~\ref{fig:prompt_n_u_nrich}.

For $\iso{U}{250}$, the emission of prompt neutrons is dominated by nuclei with $A \approx 100$ and $150$ in the moderate double peaks. This is because the TKE is relatively low, and the TXE is high. In the case of $\iso{U}{255}$, which exhibits only the single peak in the mass distribution, neutrons are mainly emitted from nuclei with $A \approx 128$, which have a relatively small number of prompt neutron emissions (Fig.~\ref{fig:prompt_n_u_nrich}) because the TKE is high and the corresponding TXE is expected to be relatively small. The calculated average number of prompt neutrons is $\langle \rm{n} \rangle = 4.185$ for $\iso{U}{250}$, which is larger than the case of $\iso{U}{236}$ with $\langle \rm{n} \rangle = 2.413$ mainly due to lower neutron binding energy. However, the average number of prompt neutrons is $\langle n \rangle = 3.434$ for $\iso{U}{255}$, which is smaller than $\iso{U}{250}$. From our calculations, neutron-rich U isotopes, which have not been experimentally identified, may exhibit a higher number of prompt neutrons than fission of nuclei along $\beta$-stability line; however, depending on fission modes and TKE distributions, as in the case of neutron-rich $\iso{U}{255}$, the number of prompt neutrons would not monotonically increase with neutron number.

\section{Summary and conclusions}\label{sec:summary}

In this study, we calculated the fission properties of uranium isotopes with a newly developed method based on a dynamical fission method (KiLM) and a HFSM (CCONE). Using an adjusted set of model parameters, we successfully reproduce experimental fission-fragment distributions, TKE, and prompt neutron emissions for the induced fission of $\iso{U}{236}$, which is a compound nucleus of ${\rm n} + \iso{U}{235}$, and two very neutron-rich uranium isotopes, i.e., $\iso{U}{250}$ and $\iso{U}{255}$, which are not experimentally confirmed, but are relevant to r-process nucleosynthesis. Our results are summarised as follows:
\begin{enumerate}

\item We accurately calculated the fission yields and TKE of $\iso{U}{236}$ using the KiLM with appropriate physical parameters, successfully reproducing the experimental values. We applied the same method to the induced fission of $\iso{U}{250}$ and $\iso{U}{255}$. These results were consistently and smoothly connected to the subsequent HFSM calculations.

\item The post-neutron emission properties of $\iso{U}{236}$ were explained from the physical point of view, and the experimental data were well reproduced. The average number of prompt neutrons, with $\langle {\rm n} \rangle = 2.574$, was in excellent agreement with the experimental value.

\item We performed neutron emission calculations of very neutron-rich uranium isotopes where experimental data are unavailable. In the case of asymmetric fission, $\iso{U}{250}$, we obtained $\langle {\rm n} \rangle = 4.185$, while the mass-symmetric fission of $\iso{U}{255}$ we had $\langle {\rm n} \rangle = 3.434$. From these results, we concluded that the number of prompt neutrons does not necessarily increase with neutron number, and it is important to understand fission mode and TKE distributions.

\end{enumerate}

Our novel calculation method, combining the Langevin and HFSM approaches, has successfully reproduced experimental data for $\iso{U}{236}$, including fission products and TKE. The remaining discrepancies can potentially be resolved by refining the model used in dynamical calculations, particularly in cases where experimental evaluations have lower accuracy. We can improve reproducibility by reducing symmetry in the nuclear shape parameters. Future advancements, building upon the findings of this study, hold the potential for enhancing our understanding of the underlying physics. Since there are other nuclei besides $\iso{U}{236}$ for which neutron emission has been measured, it will be intriguing to extend the application to other nuclei for future research.

Even in the absence of experimental data for neutron-rich nuclei, i.e., $\iso{U}{250}$ and $\iso{U}{255}$, various quantities related to fission and neutron emission can be predicted with fundamental physical validity. This region represents the transition from mass-asymmetric fission to symmetric fission as the mass number increases with neutron excess. In the range of asymmetric fission, where there is an excess of neutrons, the number of emitted neutrons tends to increase. Conversely, in the range of symmetric fission, the neutron excess tends to decrease, resulting in a decrease in the number of emitted neutrons. This behavior can be understood by examining the distribution of TKE.

In future, improving the reproducibility of experimental data will be crucial, along with the systematic development of highly accurate theoretical predictions. Understanding the nature of fission, especially in neutron-rich nuclei, is vital for applications in r-process nucleosynthesis occurring in space. The study of fission effects in neutron-rich nuclei heavily relies on theoretical approaches, which have primarily focused on half-life and fission distribution systematics. However, the results obtained from dynamical models have yet to be fully utilized in practical applications. It is important to continue investigating the effects of symmetric fission in neutron-rich nuclei and their associated neutron emission numbers, as suggested in this study, to advance our understanding of r-process nucleosynthesis.

\begin{acknowledgments} 
The authors thank O. Iwamoto for his support for using the CCONE code. The also authors acknowledge helpful discussion with participants at ``RIBF-ULIC-miniWS038,'' funded by RIBF ULOR at RIKEN. Parts of the computations shown in the present study were carried out on computer facilities on CfCA at NAOJ and YITP at Kyoto University. The project was financially supported by JSPS KAKENHI (19H00693, 20H05648, 21H01087, 21H01856, 22K20373). N.N. was supported by the RIKEN Incentive Research Projects.    
\end{acknowledgments} 


\bibliographystyle{unsrt}

\end{document}